%% file: manuscript.tex
\input{env-settings}

\begin{document}

\title{
LEAP: LLM Inference on Scalable PIM-NoC Architecture with Balanced Dataflow and Fine-Grained Parallelism
\thanks{
This work was supported by the National Research Foundation, Prime Minister’s Office, Singapore, under its Competitive Research Program (NRF-CRP24-2020-0002 and NRF-CRP24-2020-0003), and the Ministry of Education (Singapore) Academic Research Fund Tier 2 (MOE-T2EP50221-0008 and MOE-T2EP50123-0019) and Tier 1 (FY2025). 
}
\thanks{\textsuperscript{\ddag} Both authors contributed equally to this work.}
}

\author{
Yimin Wang\textsuperscript{\dag\ddag}, Yue Jiet Chong\textsuperscript{\dag\ddag}, and Xuanyao Fong\textsuperscript{\dag}\\
\textsuperscript{\dag} Department of Electrical and Computer Engineering, National University of Singapore, Singapore\\
Email: yimin.wang@u.nus.edu, jason.yj.chong@nus.edu.sg, kelvin.xy.fong@nus.edu.sg\\
}

\maketitle

\input{sec-abstract}
\input{sec-introduction}
\input{sec-preliminaries}
\input{sec-method}
\input{sec-result}
\input{sec-conclusion}

\bibliographystyle{IEEEtran}
\bibliography{mybibfile}

\end{document}

%% file: env-settings.tex
\documentclass[conference]{IEEEtran}
\IEEEoverridecommandlockouts
\usepackage{cite}
\usepackage{amsmath,amssymb,amsfonts}
\usepackage{algorithm}  
\usepackage{booktabs}
\usepackage{graphicx}
\usepackage{subfigure}
\usepackage{textcomp}
\usepackage{xcolor}
\usepackage{flushend}		
\usepackage{multirow}
\usepackage{booktabs}
\usepackage{svg}
\svgsetup{
    inkscapepath=i/svg-inkscape/
}
\svgpath{{svg/}}

\usepackage{algpseudocode}  

\usepackage{comment}

\usepackage{tikz}
\usepackage{pifont}

\newcommand{\blackcircled}[1]{%
  \tikz[baseline=(char.base)]{
    \node[shape=circle, draw, fill=black, inner sep=0.5pt] (char)
    {\color{white}{#1}};
  }%
}

\usepackage{mathtools, nccmath}
\usepackage{xpatch}
\xpatchcmd{\NCC@ignorepar}{%
\abovedisplayskip\abovedisplayshortskip}
{%
\abovedisplayskip\abovedisplayshortskip%
\belowdisplayskip\belowdisplayshortskip}
{}{}

\usepackage{cases}

\usepackage{bbm}

\usepackage{lipsum}

\usepackage{amsmath}

\def\BibTeX{{\rm B\kern-.05em{\sc i\kern-.025em b}\kern-.08em
    T\kern-.1667em\lower.7ex\hbox{E}\kern-.125emX}}

%% file: sec-abstract.tex
\begin{abstract}

Large language model (LLM) inference has been a prevalent demand in daily life and industries. 
The large tensor sizes and computing complexities in LLMs have brought challenges to memory, computing, and databus. 
This paper proposes a computation/memory/communication co-designed non-von Neumann accelerator by aggregating processing-in-memory (PIM) and computational network-on-chip (NoC), termed LEAP. 
The matrix multiplications in LLMs are assigned to PIM or NoC based on the data dynamicity to maximize data locality. 
Model partition and mapping are optimized by heuristic design space exploration. 
Dedicated fine-grained parallelism and tiling techniques enable high-throughput dataflow across the distributed resources in PIM and NoC. 
The architecture is evaluated on Llama 1B/8B/13B models and shows $\sim$2.55$\times$ throughput (tokens/sec) improvement and $\sim$71.94$\times$ energy efficiency (tokens/Joule) boost compared to the A100 GPU. 

\end{abstract}


%% file: sec-introduction.tex
\section{Introduction}

    Due to the massive data volume and computational intensity of large language models (LLMs), current hardware platforms face significant bottlenecks in memory capacity/bandwidth, compute scheduling, and hardware communication energy overhead.
    Processing-in-memory (PIM) is a widely explored design technique to accelerate AI workloads by bringing compute into the memory~\cite{NNano-2020-IMC-Review, NE-2023-full-spectrum}. 
    PIM speeds up matrix multiplication ($A\cdot B$) with a dynamic matrix $A$ and a static matrix $B$ (DSMM), which is suitable for the operations with pre-trained weights, \textit{e.g.}, the projection and fully connected layers in LLMs. 
    However, LLMs also contain immense matrix multiplications between runtime-generated dynamic matrices $A$ and $B$ (DDMM) in the attention operations. 
    These DDMMs are less suitable for traditional PIM due to high time and energy costs of dynamically reprogramming memory cells.
    Moreover, as LLMs scale in model size and input sequence length, the proportion of DDMMs increases substantially.

    To address this, existing PIM-based systems often offload DDMMs to separate computing units, including hybrid PIM arrays with configurable precision~\cite{TODAES-2024-H3D},  transposable structures~\cite{ISSCC-2022-trancim}, or fully digital accelerators~\cite{JETCAS-2024-HALO}. 
    This results in heterogeneous architectures, where computation mapping and scheduling depend heavily on the stationarity/dynamicity of the data. Such mapping challenges are further intensified in systems scaled via network-on-chip (NoC), which introduces additional design complexity and interconnect overhead.
    However, most current PIM systems only support algorithm-specific DDMMs and use custom interconnects at limited scales~\cite{ICCAD-2020-retransformer, ISSCC-2022-trancim, TCAD-2023-CPSAA}, falling short in terms of system scalability and data flow flexibility. 

    In this work, we present a hardware-software co-design approach to enable scalable and flexible acceleration of LLM inference on heterogeneous PIM architectures. 
    Our end-to-end framework provides partitioning, mapping, and scheduling of LLM inference workloads with awareness of data stationarity and system heterogeneity. 
    In addition, the hardware architecture integrates local compute and memory units within a scalable NoC to support both DDMM-specific dataflows and general aggregation operations. 
    The key contributions of this work are summarized below: 

\begin{itemize}

    \item A fine-grained model partitioning and a heuristically optimized spatial mapping strategy to achieve high PIM utilization and structured layout. 

    \item Temporal scheduling approach that incorporates dedicated context window tiling and efficient key-value caching (KV cache), ensuring balanced NoC traffic and utilization. 

    \item A custom NoC capable of efficient data communications, DDMMs, and aggregations, with re-programmability via a dedicated instruction set. 

    \item The overall system achieves $\sim{}2.55\times$ throughput improvement and $\sim{}71.94\times$ energy efficiency in the inference of the Llama model compared to the A100 GPUs. 

\end{itemize}

%% file: sec-preliminaries.tex
\section{Preliminaries}

\begin{figure}[t]
\centerline{\includegraphics[width=90mm]{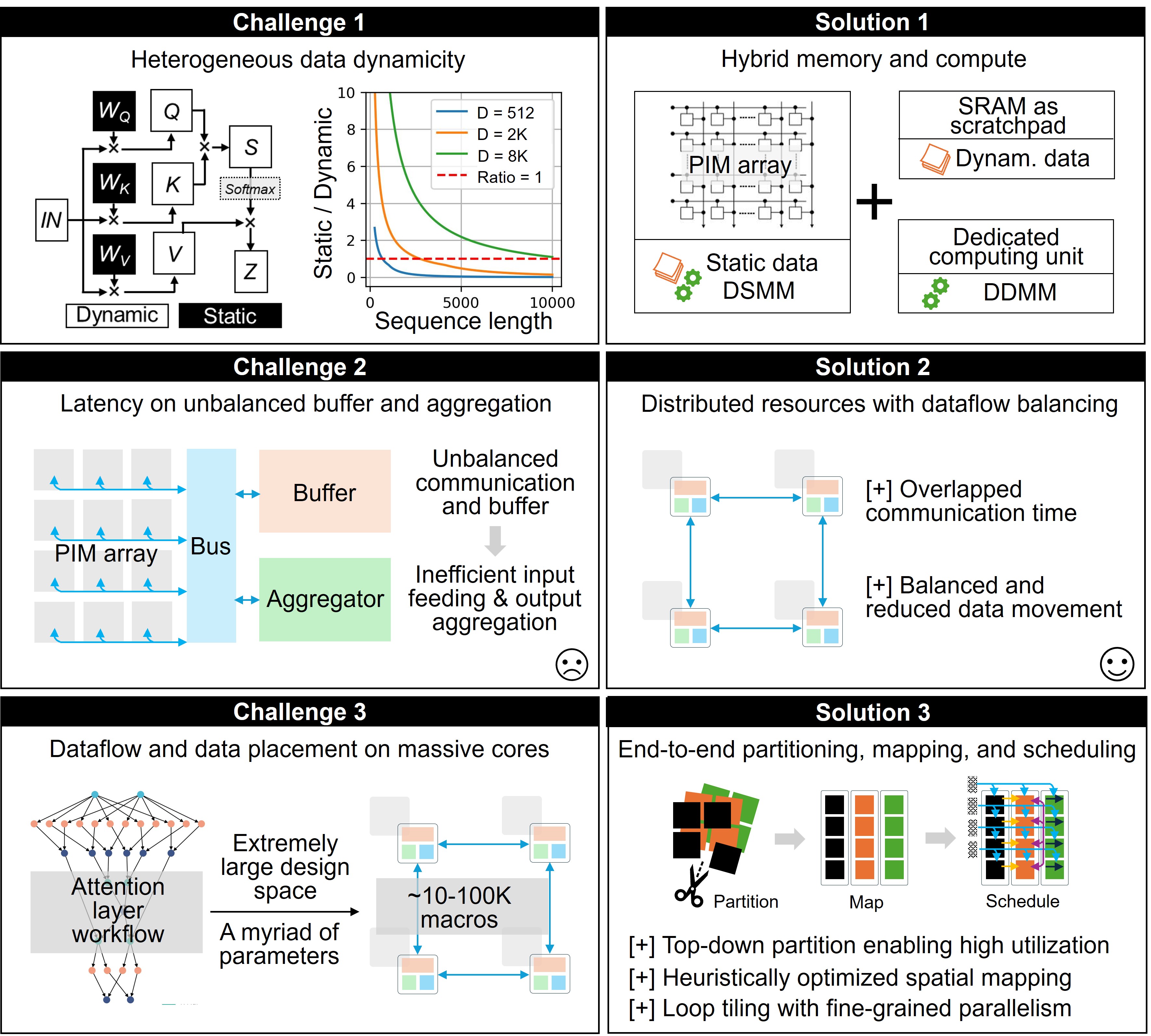}}
\caption{
Design challenges and solutions in accelerating LLM inference. 
}
\label{fig-design-challenge}
\end{figure}

\subsection{Data Stationarity in LLMs}

    Recent commercial LLMs, such as GPT~\cite{NeurIPS-2020-GPT3} and the Llama series~\cite{Meta-2023-llama2, Meta-2024-llama3, Meta-2024-llama3.1, Meta-2024-llama3.2}, are predominantly decoder-only Transformers. 
    Each decoder layer comprises attention and feed-forward sublayers, which involve successive matrix multiplications (MMs) and matrix-vector multiplications (MVMs). 
    Although these models rely on static pre-trained weights, the attention mechanism generates significant dynamic data during inference. 
    To quantify this, consider an attention layer with embedding dimension $D$ and sequence length $S$. 
    The amount of static data (pre-trained weights) is: 
    \begin{equation}
        \begin{split}
            DA_{static} = 4D^2
        \end{split}
    \end{equation}
    which is independent of the input sequence length. 
    The dynamic data generated at runtime is: 
    \begin{equation}
        \begin{split}
            DA_{dynamic} = 5SD + S^2
        \end{split}
    \end{equation}
    which correlates to $S$. 
    As $S$ increases, the ratio of static to dynamic data decreases: 
    \begin{equation}
        \begin{split}
            \frac{DA_{static}}{DA_{dynamic}} = \frac{4D^2}{5SD + S^2} \overset{S=D}{=}\frac{2}{3}
        \end{split}
        \label{eq-ising-model}
    \end{equation}
    In real-world applications, where $S \gg D$, dynamic data increasingly dominates, particularly under the high-demand sequence length scaling. 
    This insight arrives at the \textbf{Challenge 1}: the heterogeneous nature of data in LLMs necessitates differentiated compute and memory strategies for static and dynamic data.

\subsection{PIM Scaling-up}

    PIM accelerates MMs/MVMs with static weights, \textit{e.g.}, DSMMs, by performing computation within non-volatile memory. 
    However, the typical array size is limited to $32\sim256$~\cite{TCAD-2018-NeuSimV1, TCAD-2020-NeuSimV2, TCASI-2019-optimizing-mapping}, making large-scale operations reliant on partitioning across many arrays. 
    This introduces significant overhead in buffering and aggregating partial results, which greatly diminishes overall efficiency~\cite{VLSI-2024-Systolic-CIM-Zigzag} if the shared buffer and aggregators are allocated in an unbalanced manner, as shown in Fig.~\ref{fig-design-challenge}. 
    Therefore, \textbf{Challenge 2} is that scaling PIM-based MMs/MVMs requires efficient interconnection and aggregation mechanisms to mitigate performance bottlenecks.

\begin{figure}[t]
\centerline{\includegraphics[width=80mm]{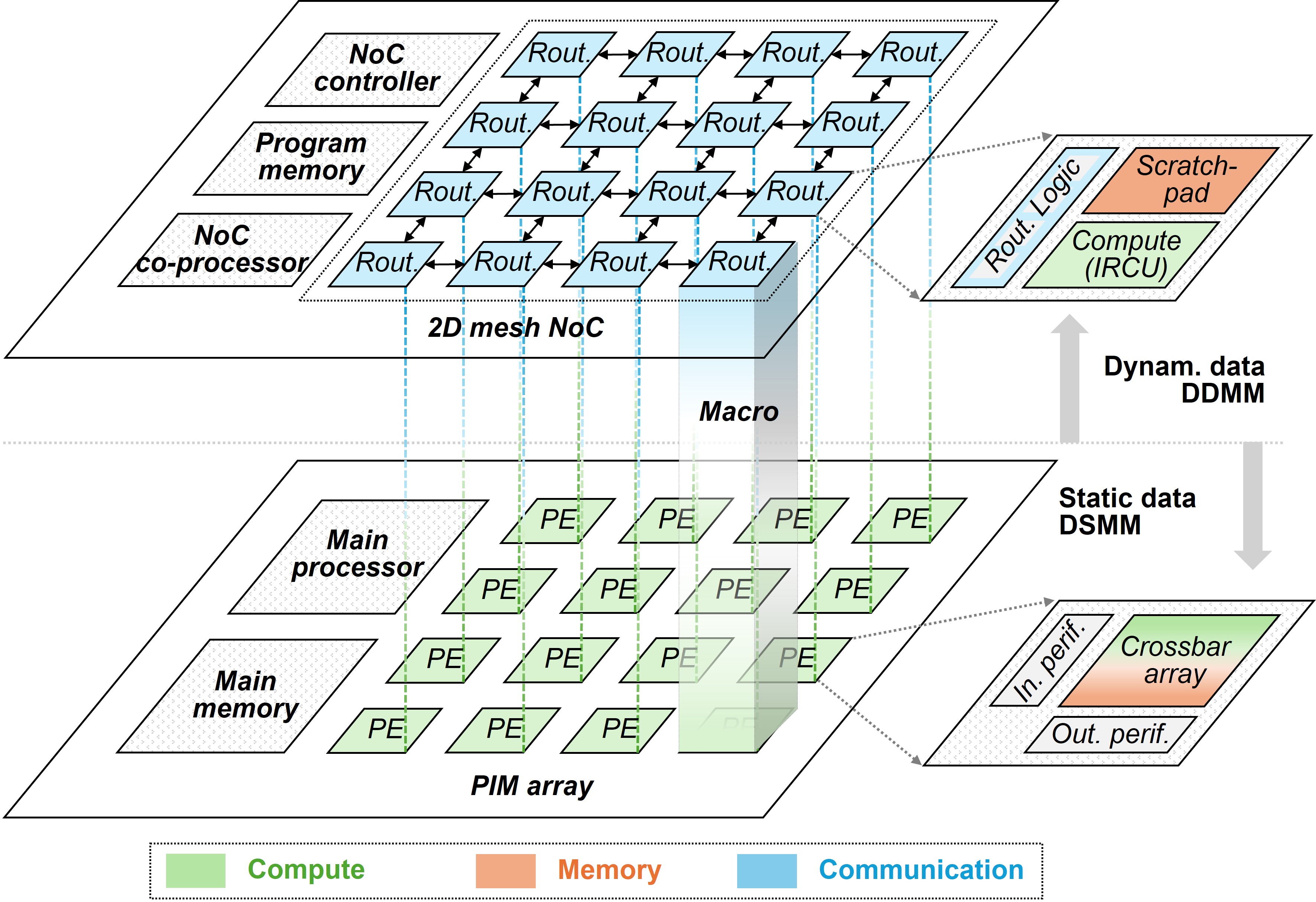}}
\caption{
The proposed aggregated PIM-NoC architecture with distributed fine-grained compute-memory-communication resources. 
}
\label{fig-pim-noc-hardware}
\end{figure}

\subsection{Target Architecture}

    To address these challenges, we target a hybrid architecture that combines PIM with a scalable NoC, referred to as aggregated PIM-NoC. 
    The architecture integrates: 
    i) PIM processing elements (PEs) -- the non-volatile memory units capable of in-place DSMM computations; and 
    ii) computational routers -- the dedicated computing units, termed in-router computing units (IRCUs) and SRAM-based scratchpad, optimized for DDMMs and partial results aggregation, as shown in Fig.~\ref{fig-pim-noc-hardware}. 
    Each router-PE pair forms a macro, the basic building block of a distributed 2D mesh system with unified compute, memory, and communication resources. 
    Given these novel features compared to traditional von Neumann architecture, the \textbf{Challenge 3} is that efficiently partitioning, mapping, and scheduling LLM workloads on such a spatially distributed and heterogeneous architecture demands compilation framework innovations due to the vast design space. 
    In this work, we demonstrate an end-to-end framework that systematically addresses these challenges. 

%% file: sec-method.tex
\section{Model Partitioning and Spatial Mapping}

This section introduces the partitioning scheme for the projection weight matrices and a spatial mapping strategy that deploys the partitioned matrices onto the PIM PEs.

\begin{figure*}[t]
\centering

\subfigure[]{
\begin{minipage}[t]{0.76\linewidth}
\centering
\includegraphics[width=138mm]{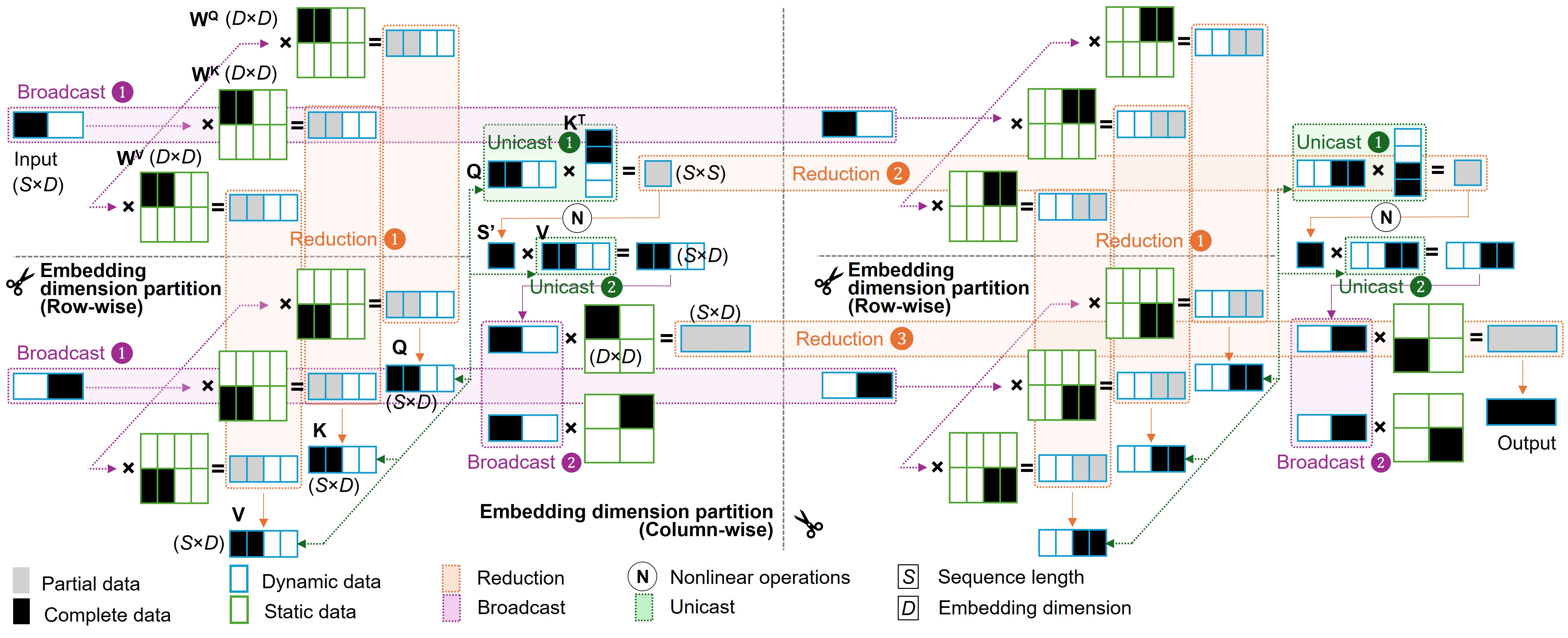}
\end{minipage}%
}%
\subfigure[]{
\begin{minipage}[t]{0.24\linewidth}
\centering
\includegraphics[width=45mm]{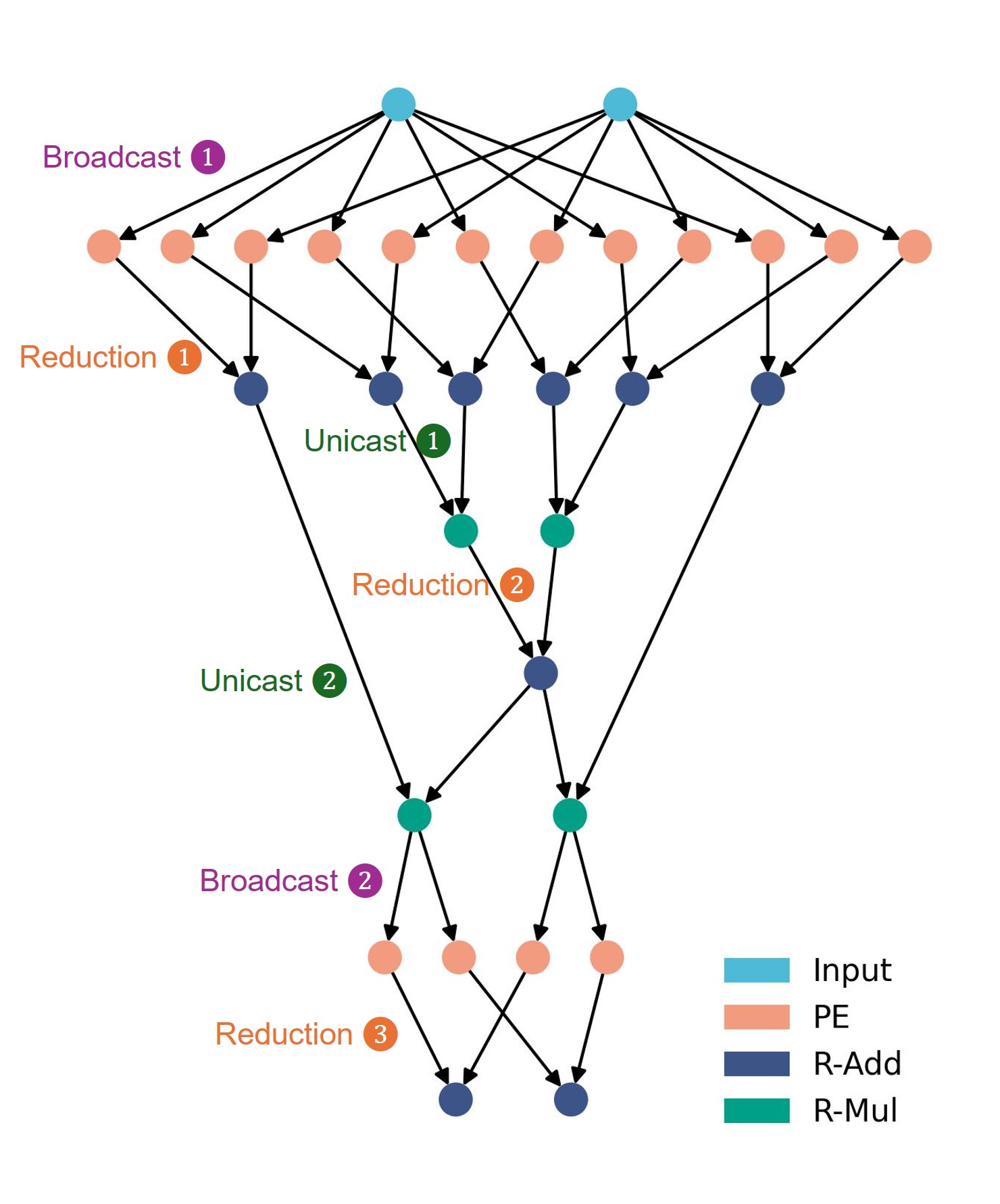}
\end{minipage}%
}%

\centering
\caption{
(a)
Partitioning of an attention layer. 
This illustration considers multi-head attention (MHA), whereas other attention variants like group-query attention (GQA) can degrade to this scheme by matrix duplication accordingly. 
(b)
DAG represents the data and operations in the partitioned attention layer. 
``R-Add" and ``R-Mul" are short for addition and multiplication operations in routers. 
}
\label{fig-partition-overview}
\end{figure*}

\begin{figure}[t]
\centerline{\includegraphics[width=85mm]{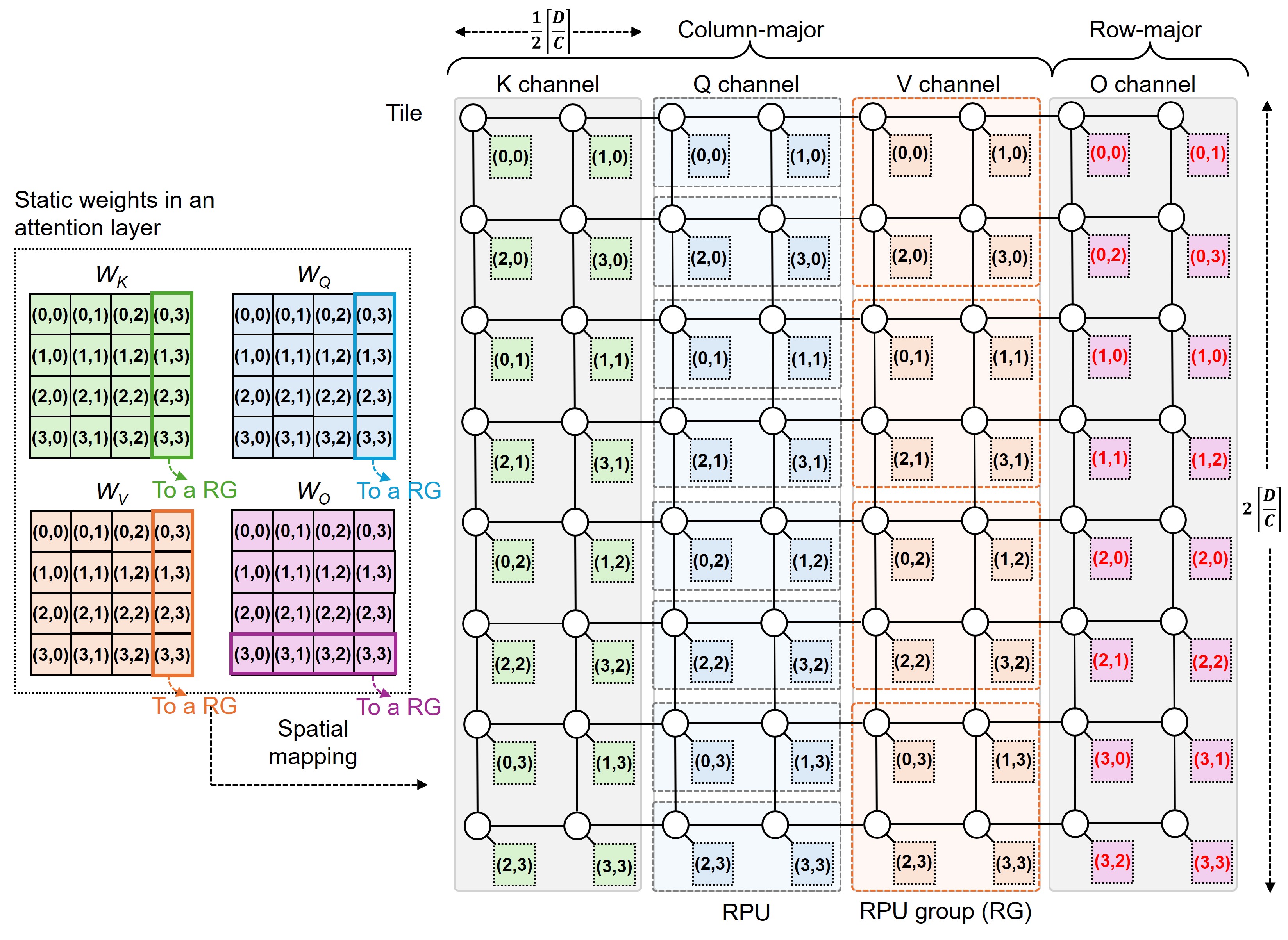}}
\caption{
The spatial mapping used in this work. 
Static weight matrices are mapped spatially across the crossbar arrays in PEs. 
}
\label{fig-weight-mapping}
\end{figure}

\subsection{Partitioning}

    Partitioning is applied along both row and column dimensions of the static weight matrices, $\mathbf{W_Q}$, $\mathbf{W_K}$, $\mathbf{W_V}$, and $\mathbf{W_O}$ $\in \mathbb{R}^{D\times D}$, to fit the dimensions of the crossbar arrays, as illustrated in Fig.~\ref{fig-partition-overview}~(a) using an attention layer as an example. 
    The number of crossbar arrays required to store each matrix after partitioning is $\lceil \frac{D}{C} \rceil^2$, where $C$ denotes the width and height of a crossbar array. 
    Intermediate data such as $\mathbf{Q}$, $\mathbf{K}$, $\mathbf{V}$, and $\mathbf{S}$ are also partitioned, introducing additional collective communication steps for broadcasting partitioned inputs (Broadcast \blackcircled{1}/\blackcircled{2}) or reducing partial outputs (Reduction \blackcircled{1}/\blackcircled{2}/\blackcircled{3}). 
    The data dependencies and communication requirements among partitioned matrices and operations are represented by the directed acyclic graph (DAG) $\mathcal{G}$ shown in Fig.~\ref{fig-partition-overview}~(b). 

    To execute the whole attention layer on the PIM-NoC architecture, all nodes and edges in $\mathcal{G}$ must be mapped onto either PEs or routers. 
    This involves two main steps: 
    (i) spatial mapping, which assigns partitioned static weights and their associated DSMM operations (represented as orange nodes) in $\mathcal{G}$ to the crossbar arrays in PEs, and 
    (ii) temporal mapping, which schedules the storage of intermediate data in scratchpads, assigns DDMM operations to IRCUs, and orchestrates the temporal dataflow across the NoC.

\subsection{Spatial Mapping on PEs}

    A na\"ive approach to achieving optimal spatial mapping is an exhaustive design space exploration, which is computationally prohibitive due to the vast number of possible mappings. 
    For instance, a static weight matrix of size $1024 \times 1024$ can be partitioned into 64 sub-matrices, each fitting a $128 \times 128$ crossbar array. 
    The total number of possible mappings is $^{64}P_{64} \approx 1.27 \times 10^{89}$, leading to an extremely large and impractical search space. 
    To efficiently obtain a near-optimal mapping, we introduce the following heuristic constraints: 
    \begin{itemize}
        \item Sub-matrices originating from the same weight matrix must be placed within a spatially proximate region. 
        \item This region should have a rectangular shape to facilitate regular dataflows and reduce routing complexity. 
        \item The sub-matrices within this region should be ordered in a row-major or column-major fashion. 
    \end{itemize}
    These constraints dramatically reduce the number of mapping candidates by approximately $10^{86}\times$, resulting in only 1440 valid configurations. 
    We define the cost function for spatial mapping as the total communication time, $\mathbb{C} = T_{\text{comm}}^{\text{tot}}$, and use a naïve X–Y routing algorithm as the baseline for communication cost estimation.
    With the constrained search space, the spatial mapping exploration completes within 20 seconds. 
    The selected spatial mapping strategy is visualized in Fig.~\ref{fig-weight-mapping}, and its optimality will be evaluated in Section VI. 
    The entire attention layer is mapped onto a square region comprising $2\lceil \frac{D}{C} \rceil \times 2\lceil \frac{D}{C} \rceil$ macros, referred to as a tile. 
    Each individual projection weight matrix is allocated to a rectangular region of $2\lceil \frac{D}{C} \rceil \times \frac{1}{2} \lceil \frac{D}{C} \rceil$ macros within this space, referred to as a channel. 
    Sub-matrices from $\mathbf{W_Q}$/$\mathbf{W_K}$/$\mathbf{W_V}$ are mapped in a column-major, while those from $\mathbf{W_O}$ are mapped in a row-major. 
    We define the following terminology used throughout the rest of the paper: 
    i) A row-wise processing unit (RPU) refers to a single row of macros within a channel; and 
    ii) An RPU group (RG) consists of the RPUs that store a column-wise partition of $\mathbf{W_Q}$/$\mathbf{W_K}$/$\mathbf{W_V}$ or a row-wise partition of $\mathbf{W_O}$ as denoted in Fig.~\ref{fig-weight-mapping}.

\begin{figure*}[t]
\centerline{\includegraphics[width=170mm]{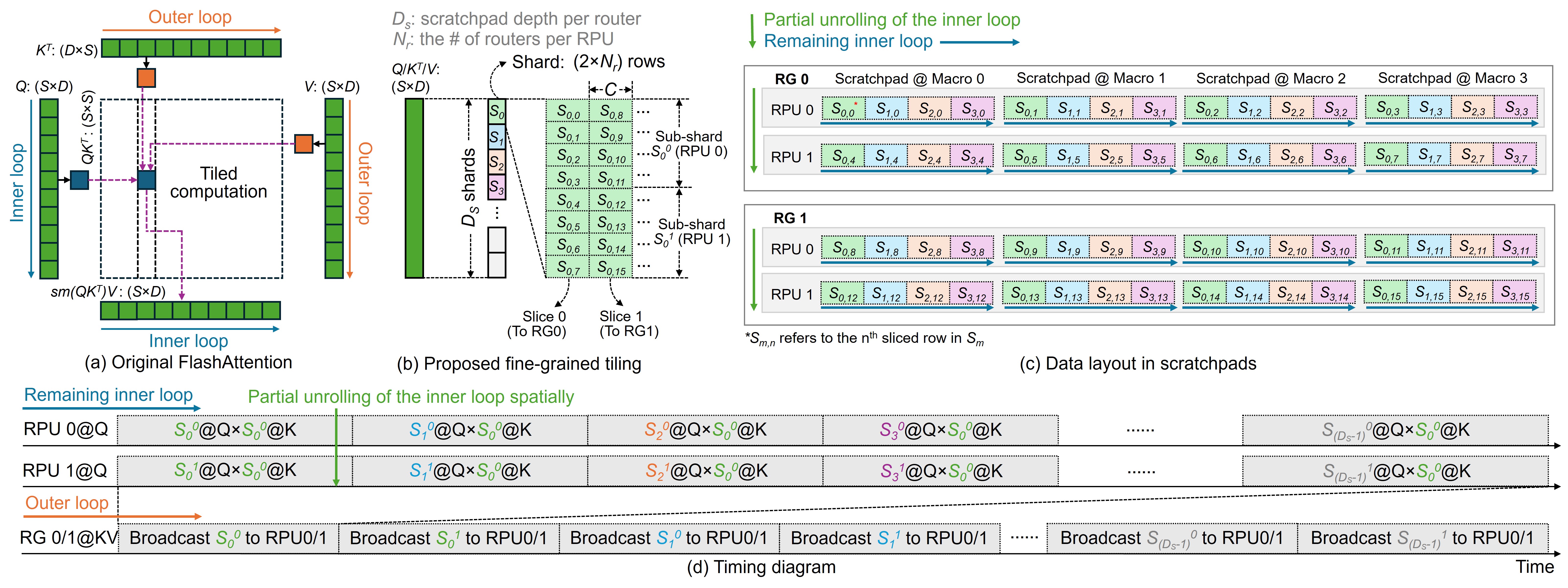}}
\caption{
Context window tiling. 
(a) Original FlashAttention. 
(b) $Q$/$K$/$V$ tiling. 
(c) Data layout in scratchpads. 
(d) Timing diagram. 
}
\label{fig-flashattention-dataflow}
\end{figure*}

\begin{figure*}[t]
\centerline{\includegraphics[width=180mm]{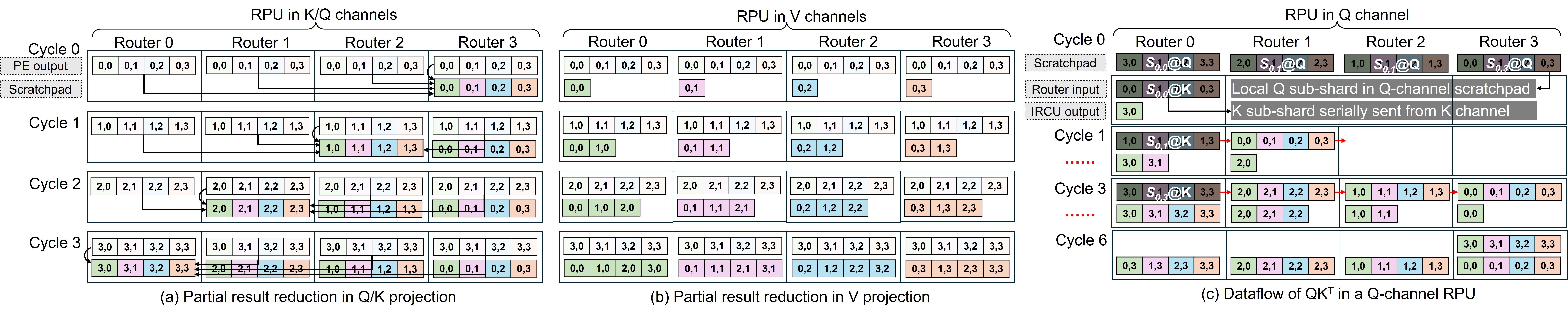}}
\caption{
Dataflow in RPUs when processing a shard by demystifying on a 4$\times$4 matrix. 
(a) Row-major reduction of $Q$/$K$. 
(b) Column-major reduction of $V$. 
(c) Dataflow of $\mathbf{Q}\mathbf{K}^T$ operation in prefill stage. 
}
\label{fig-fused-dataflow}
\end{figure*}

\section{Dataflow in Temporal Mapping}

    This section introduces the temporal mapping, which involves storing dynamic data in scratchpads and managing DDMMs on IRCUs. 
    While spatial mapping addresses partitioning along the embedding dimension, partitioning along the context window (\textit{i.e.}, token sequence length) is achieved through loop tiling in the temporal mapping stage. 
    LLM inference typically consists of two computing phases: 
    i) the prefill stage, which is MM-intensive and processes all input tokens in batch, and 
    ii) the decode stage, which is MVM-intensive and generates output tokens one at a time. 
    The dataflow strategies for both the prefill and decode stages are introduced in the following subsections.

\subsection{Context Window Tiling}
    
    The attention score $\mathbf{S}\in\mathbb{R}^{S\times S}$ computed as $\mathbf{Q}\mathbf{K}^T$, has a size quadratic in the context window length, which often exceeds on-chip memory capacity. 
    In GPU-based scenarios, FlashAttention~\cite{NeurIPS-2022-FlashAttention}, a loop-tiling method, is widely adopted to reduce the off-chip memory access and avoid materializing the full $\mathbf{S}$ matrix at once. 
    FlashAttention tiles $\mathbf{Q}$/$\mathbf{K}$/$\mathbf{V}$ matrices along the sequence length dimension ($S$), using two nested loops: the outer loop iterates over tiled $\mathbf{K}$/$\mathbf{V}$ matrices, while the inner loop processes the tiled $\mathbf{Q}$ matrices as illustrated in Fig.~\ref{fig-flashattention-dataflow}~(a). 

    We adopt this nested loop structure in our design, but introduce three key distinctions: 
    \textbf{(i)} A dedicated fine-grained tiling scheme is applied to $\mathbf{Q}$/$\mathbf{K}$/$\mathbf{V}$. 
    These matrices are partitioned into shards along two dimensions, as shown in Fig.~\ref{fig-flashattention-dataflow}~(b). 
    Each row of a shard is distributed across different routers within a RG, as illustrated in Fig.~\ref{fig-flashattention-dataflow}(c). 
    The capacity of each shard is $C_S = 2 \cdot N_r = \lceil \frac{D}{C} \rceil$, where $N_r = \frac{1}{2} \lceil \frac{D}{C} \rceil$ is the number of routers in an RPU. 
    With this scheme, the context window length supported by a tile is $D_S \cdot C_S$, where $D_S$ is the scratchpad depth per router. 
    \textbf{(ii)} The inner loop is spatially unrolled across the RPUs, exploiting their parallelism to improve throughput. 
    \textbf{(iii)} The outer loop is implemented by rotational broadcasting of the $\mathbf{K}$/$\mathbf{V}$ shards across the RPUs within each RG as shown in Fig.~\ref{fig-flashattention-dataflow}(d). 
    The detailed dataflow for processing a single shard is described in the following subsections.

\subsection{Prefill Dataflow}

    \subsubsection{DSMM}
    During the projection step, input activations are fed from the leftmost column into Q/K/V channels (Broadcast \blackcircled{1} as in Fig.~\ref{fig-partition-overview}(b)). 
    Each PE within the Q/K/V channels generates a vector of partial results per cycle. 
    These partial results are then aggregated within each RG (Reduction \blackcircled{1}). 
    It is noted that the aggregation sequence differs across channels: row-major in the K/Q channels and column-major in the V channel, as shown in Fig.~\ref{fig-fused-dataflow}~(a) and (b), respectively. 
    The aggregated results are stored in the scratchpad according to the layout strategy introduced in the previous subsection. 

    \subsubsection{DDMM}

    Each shard $\mathbf{K}^s$ within the K-channel RPU is read from the scratchpad and transmitted rightward to the corresponding Q-channel RPU within the same row, whose pipelining is shown in Fig.~\ref{fig-fused-dataflow} (Unicast \blackcircled{1}). 
    Each Q-channel RPU computes a local partial attention score, which is then aggregated through a vertical reduction across Q-channel RGs to obtain the full attention score shard $\mathbf{S}^s$ (Reduction \blackcircled{2}). 
    A local Softmax operation is applied as $\mathbf{S}^s$ is passed to the V channel. 
    We adopt the softmax algorithm from FlashAttention, which requires storing intermediate values such as $\mathbf{O}^s$ and \texttt{rowmax}, etc. 
    These are held in the O-channel scratchpad. 
    Partial results received from the V channel are combined element-wise with previously accumulated values and written back to the O-channel scratchpad (Unicast \blackcircled{2}). 
    Once completed, each full $\mathbf{O}^s$ shard is broadcast across the corresponding O-channel RG (Broadcast \blackcircled{2}), followed by a vertical reduction to finalize the output (Reduction \blackcircled{3}).

    Under the proposed spatial and temporal mapping strategy, dataflow aligns cleanly along horizontal and vertical paths. 
    This regularity enables an expected balance between minimizing traffic collisions and maximizing parallelism.

\subsection{Decode Dataflow}

    In the decode stage, two key differences distinguish it from the prefill stage:
    \textbf{(i)} only a single newly-generated Q vector is involved in the attention computation, and \textbf{(ii)} newly-generated K/V vectors are incrementally appended into the scratchpad at each timestep. 
    Due to this limited parallelism, the $\mathbf{Q}\mathbf{K}^T$ pipeline shown in Fig.~\ref{fig-fused-dataflow} may be underutilized, leading to reduced throughput compared to the prefill stage, as will be demonstrated in Section VI. 
    Nevertheless, the caching of newly generated $\mathbf{K}/\mathbf{V}$ vectors adheres to the same placement strategy shown in Fig.~\ref{fig-flashattention-dataflow}(b), which inherently ensures balanced scratchpad utilization across routers. 
    This approach eliminates the need for additional data movement or shifting, offering an improvement over prior KV-cache management techniques such as those in \cite{AX-2025-WaferLLM}, especially on scalable architectures.

\section{Hardware Implementation}

    A dedicated NoC is designed to facilitate effective communication among PEs and to support the DDMMs and aggregations in the IRCU. 
    The NoC architecture comprises three key components: the program memory, the main controller, and the router mesh. 

\begin{figure}[t]
\centerline{\includegraphics[width=90mm]{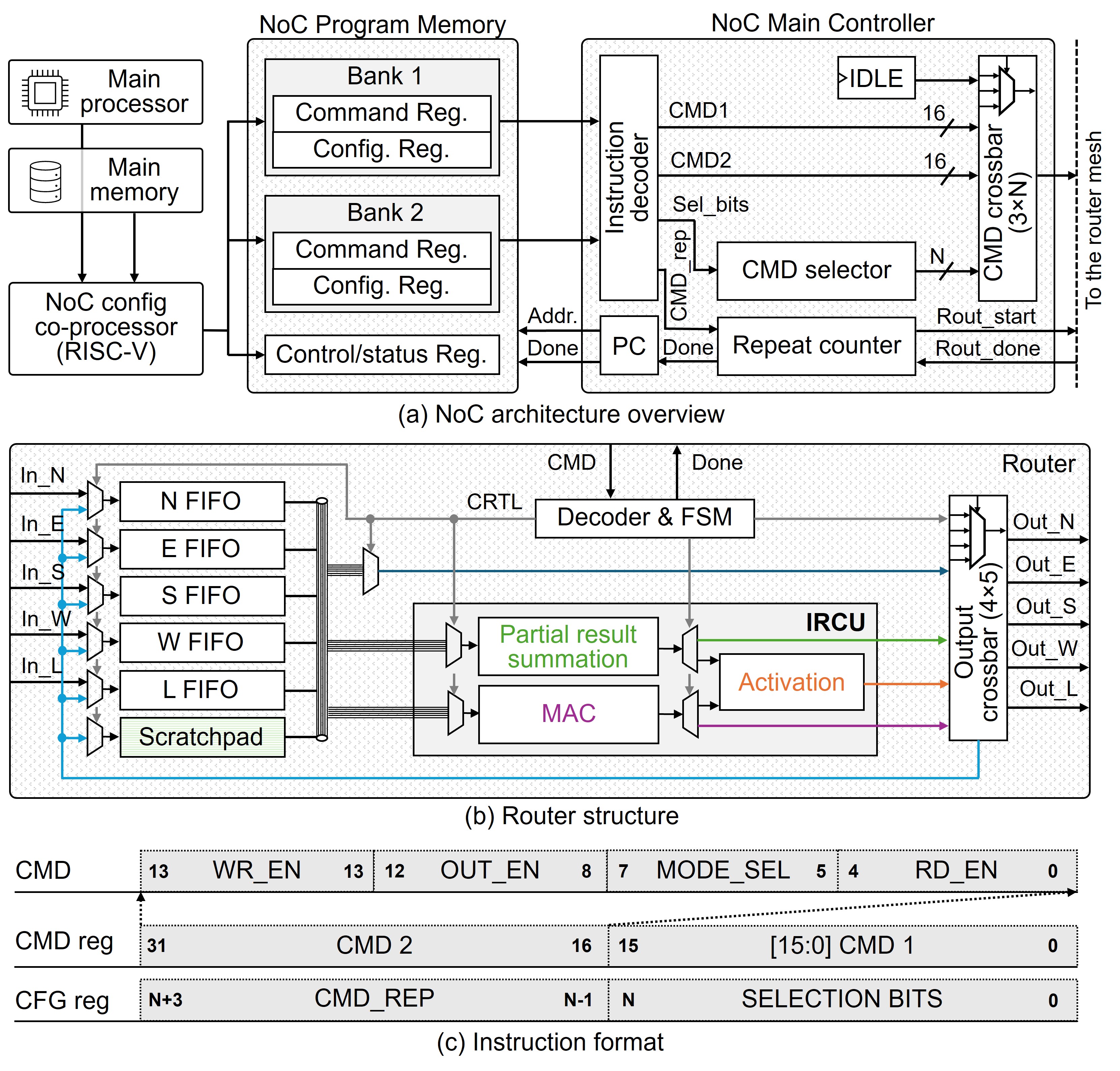}}
\caption{
Overview of the NoC system architecture. }
\label{fig-noc-architecture}
\end{figure}

    \subsection{Instruction Format}

    The co-processor programs the NoC program memory (NPM) with instructions. 
    Each instruction comprises two components: a command pair (CMD1, CMD2) and a configuration word, which are written to the command register and configuration register, respectively, as illustrated in Fig.~\ref{fig-noc-architecture}. 
    The configuration word encodes the command repetition count (CMD\_rep) and the router selection bits (Sel\_bits). 
    CMD1 and CMD2 can be executed concurrently, each directing data along a distinct, non-conflicting path. 
    This design aligns with the earlier elaborated dataflow, which shows that concurrent data movement typically occurs in at most two directions. 

    An alternative instruction-reading scheme combined with a double-bank design is employed to minimize idle time. 
    NPM consists of two independent banks, each containing a set of command and configuration registers. 
    These banks are configured alternatively by the co-processor: while the controller reads instructions from one bank, the co-processor programs the other. 
    For example, when the controller is reading from Bank 2, Bank 1 is simultaneously configured, and vice versa. 

    The NoC main controller (NMC) is responsible for fetching and decoding the instructions from the NPM to orchestrate data movement. 
    During decoding, the instruction is split into two commands, CMD1 and CMD2, which are dispatched to the router command crossbar. 
    The command crossbar is a 3-input, $N$-output structure, where $N$ corresponds to the number of routers in the network. 
    Each router concurrently executes either CMD1, CMD2, or remains IDLE, repeating the operation for the number of times specified by CMD\_rep in the configuration word. 
    A command repeat counter keeps track of the remaining repetitions by decrementing on each cycle. 
    Once the counter reaches zero, it signals the program counter (PC) to advance to the next instruction. 

    A Python API is provided to facilitate programming the LLM inference dataflow to the 2D mesh NoC. 
    The compiler then translates the user's Python code into a corresponding hex file that can be loaded into the NPM for execution.

    \subsection{Router Implementation}

    Each router includes five data I/O ports: four for interconnection with adjacent routers (North, East, South, and West), and one for communication with the locally attached PE. 
    Incoming data from these ports are buffered in dedicated FIFOs. 
    The IRCU supports key operations including: partial result summation (used in Reductions \blackcircled{1}/\blackcircled{2}/\blackcircled{3}), activation functions (\textit{e.g.}, Softmax), and multiply-accumulate (MAC) computations (used in DDMMs). 
    The output crossbar switch is a 4-input-5-output crossbar, allowing data to be routed to adjacent routers or the local PE. 
    This architecture supports multi-cast, enabling a single data packet to be forwarded to up to five destinations concurrently. 

%% file: sec-result.tex
\section{Results and Discussions}

\subsection{Experimental Setup}

    \input{tables/table_system_setup}

    \textbf{Hardware testbed}: 
    The hardware configurations are summarized in Table~\ref{table-hardware-configuration}. 
    The digital components (routers and controllers) are implemented in Verilog HDL and synthesized using Synopsys Design Compiler at a 45 nm technology node~\cite{FreePDK}. 
    Power estimation is performed with Synopsys PrimeTime, using switching activity data from post-synthesis simulations, and place-and-route is carried out using Cadence Innovus. 
    Scratchpad area and power are estimated via CACTI~\cite{CACTI}. 
    The area and power of the PIM PE, featuring a 128$\times$128 RRAM crossbar array, are adopted from \cite{TCASI-2019-optimizing-mapping}. 

    \textbf{Performace benchmark}: 
    End-to-end throughput is evaluated on various LLMs: Llama 3.2-1B~\cite{Meta-2024-llama3.2}, Llama 3-8B~\cite{Meta-2024-llama3}, and Llama 2-13B~\cite{Meta-2023-llama2}, using an instruction-level simulator customized for the proposed NoC instruction set.

\subsection{Mapping Space Exploration}

\begin{figure}[t]
\centerline{\includegraphics[width=85mm]{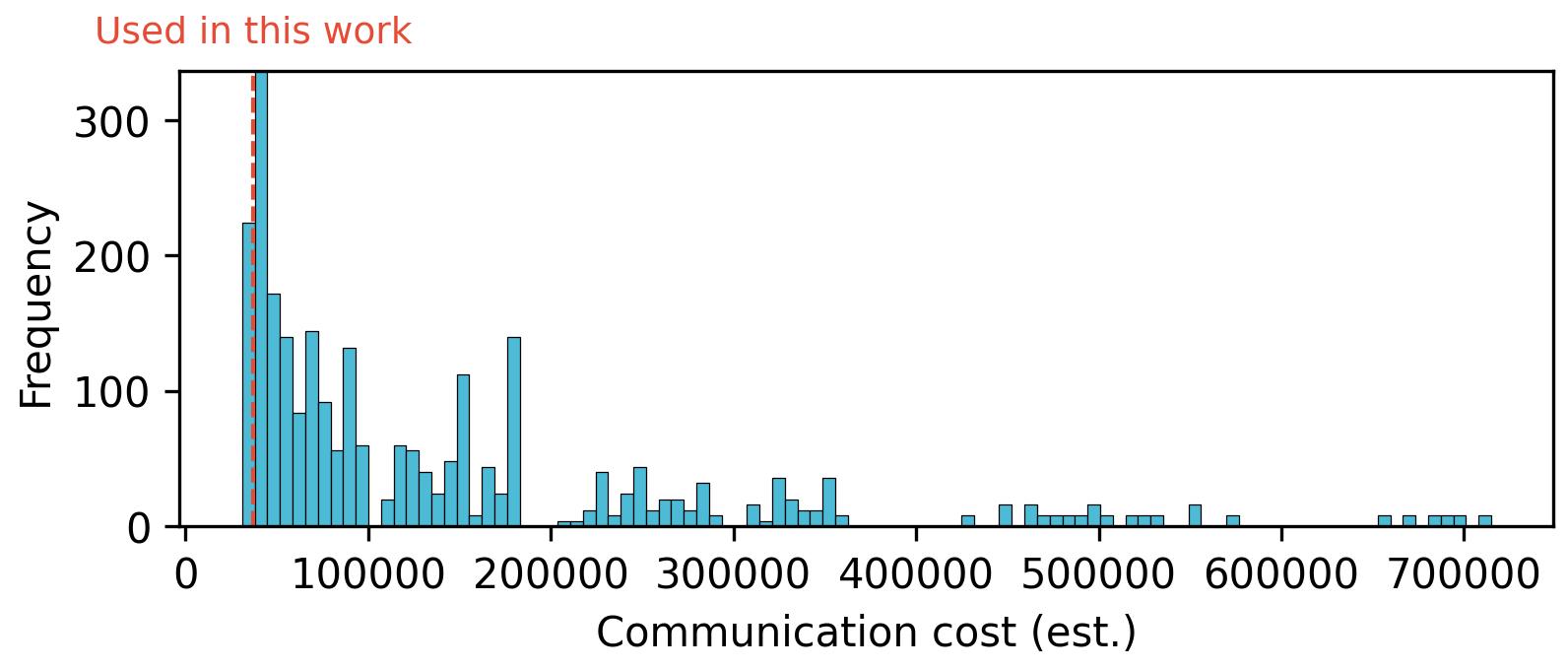}}
\caption{
Distribution of the communication cost in a spatial mapping space exploration of mapping an attention layer in Llama 3.2-1B to 1024 macros. 
}
\label{fig-rough-mapping-histgram}
\end{figure}

\begin{figure}[t]
\centerline{\includegraphics[width=65mm]{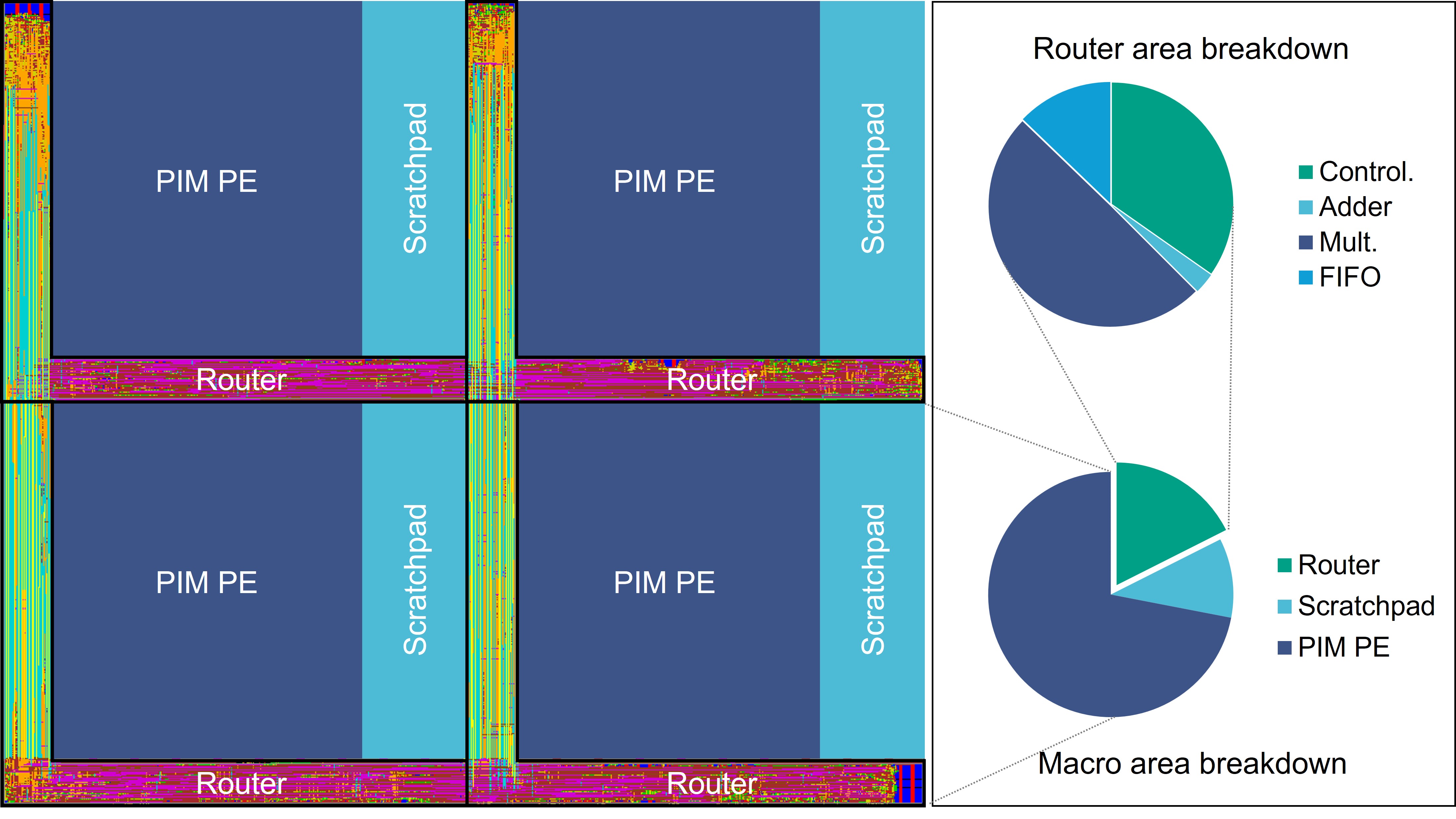}}
\caption{
Example layout of a 2$\times$2 macro array and area breakdown on the macro-level and router level. 
}
\label{fig-router-layout}
\end{figure}

    \input{tables/table_hardware_evaluation_7nm}
    \input{tables/table_throughput_comparison_hardware}

    To evaluate the optimality of the proposed spatial mapping strategy, we perform a design space exploration based on the heuristics described in Section III-B. 
    Fig.~\ref{fig-rough-mapping-histgram} shows the distribution of the communication cost for mapping an attention layer of Llama 3.2-1B onto 1024 macros, across 2,592 evaluated spatial mapping candidates. 
    The results confirm that the adopted strategy yields one of the lowest communication costs among all evaluated mappings. 
    It is worth noting that this communication cost evaluation is based on a coarse-grained X-Y routing algorithm and does not incorporate the fine-grained temporal mapping strategies discussed earlier. 
    This explains why the selected mapping, while near-optimal, is not the absolute minimum in the distribution.

\subsection{Power and Area Breakdown}

    The power and area breakdown of a macro, scaled to the 7 nm technology node, is shown in Table~\ref{table-macro-breakdown}. 
    Further breakdowns at both the macro and router levels are illustrated in Fig.~\ref{fig-router-layout}. 
    Although the router accounts for only 17.78\% of the macro’s area, it contributes to 75.10\% of the energy consumption due to its central role in data communication and dynamic processing. 
    Thanks to the scalability of the 2D mesh topology, the area distribution remains consistent even as the system scales. 

    Table~\ref{table-gpu-comparison} compares the proposed system with state-of-the-art GPUs, A100 and H100, in terms of throughput, power, and energy efficiency. 
    The throughput is evaluated with a full context window of 2048 tokens: 1024 input tokens and 1024 output tokens. 
    Compared to A100, the proposed system achieves $\sim$2.55$\times$ higher throughput and $\sim$71.94$\times$ higher energy efficiency. 
    While its throughput is lower than that of the H100, it still delivers a $\sim$24.22$\times$ improvement in energy efficiency. 
    The significant improvement in energy efficiency is attributed to the reduced data movement overhead enabled by the fully distributed compute/memory architecture and its highly optimized dataflow, in contrast to the conventional shared-memory design of GPUs.

\subsection{End-to-end Throughput}

\begin{figure}[t]
\centerline{\includegraphics[width=83mm]{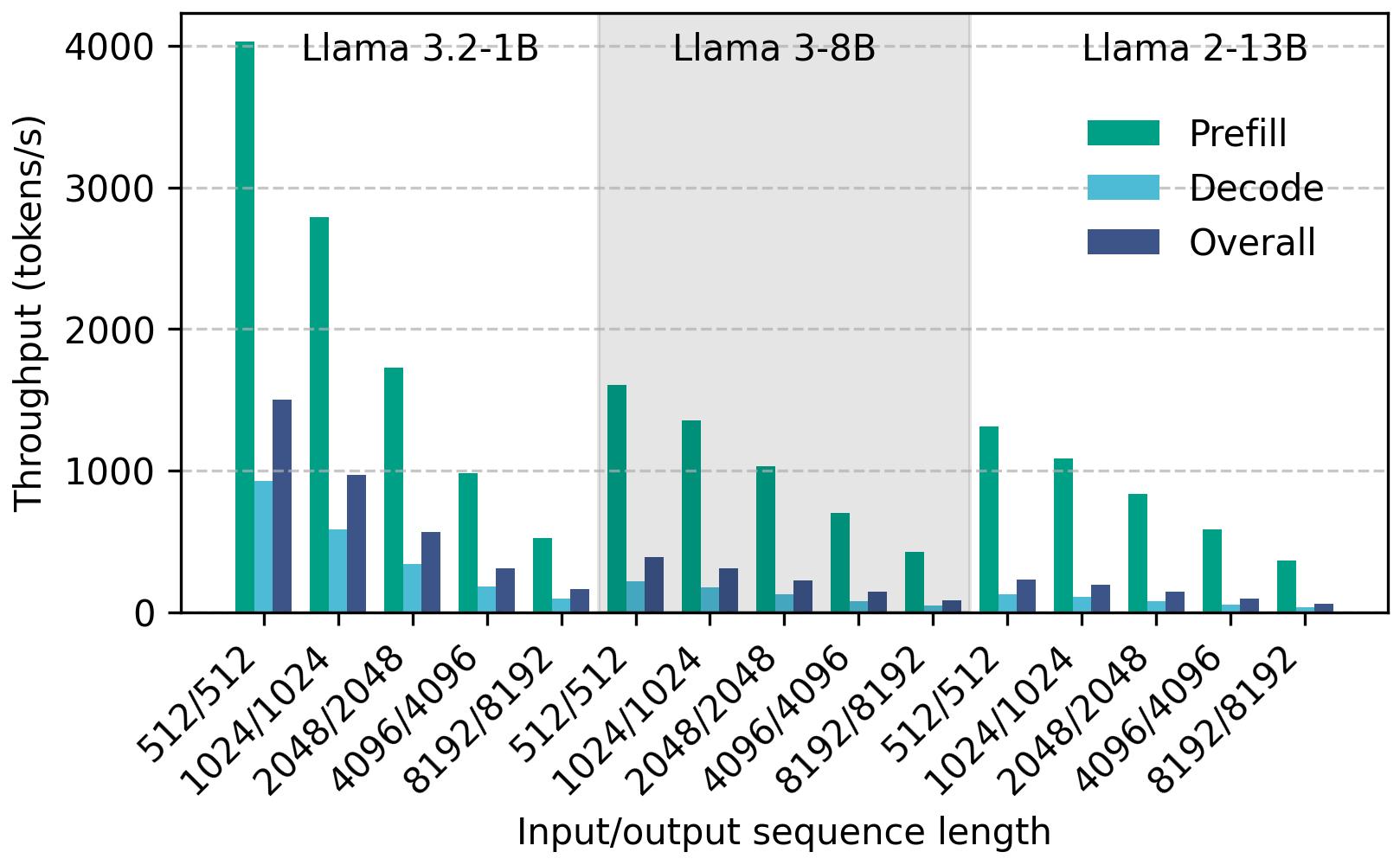}}
\caption{
Throughput under various models and input/output sequence lengths. 
}
\label{fig-throughput-model-length}
\end{figure}

\begin{figure}[t]
\centerline{\includegraphics[width=85mm]{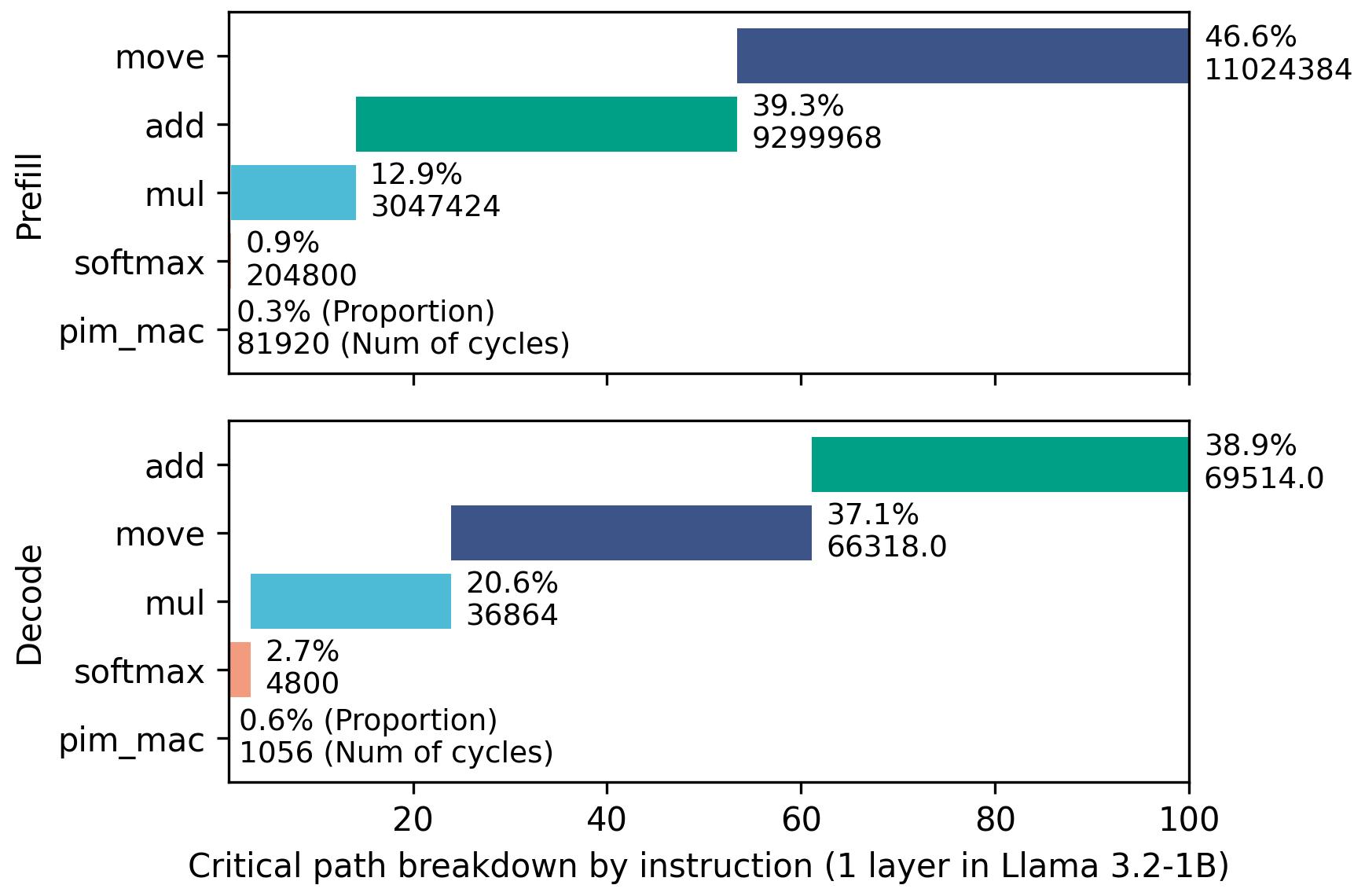}}
\caption{
Breakdown of clock cycles on the critical path by instructions in processing an attention layer and its subsequent MLP in Llama 3.2-1B. 
\texttt{mul} and \texttt{add} stands for the computations in IRCU. 
}
\label{fig-instruction-breakdown}
\end{figure}

\begin{figure}[t]
\centerline{\includegraphics[width=83mm]{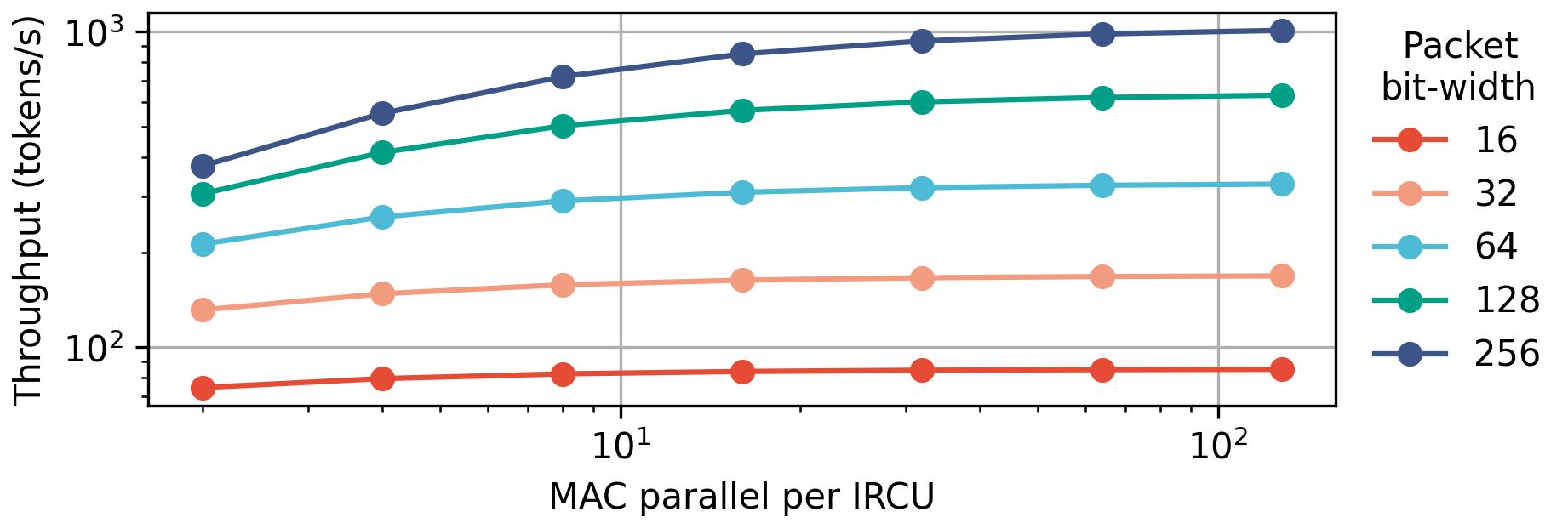}}
\caption{
Trend projection of throughput under increased packet bit-width and IRCU parallelism. 
}
\label{fig-throughput-roofline}
\end{figure}

    Fig.~\ref{fig-throughput-model-length} illustrates the inference throughput across various models and context window sizes, with further throughput breakdown into the prefill state and the decode stage. 
    The decode throughput is generally 4$\sim$6$\times$ less than that of the prefill stage. 
    This degradation is primarily due to two factors: 
    i) The number of past tokens that each newly generated token must attend to keeps increasing; and 
    ii) Only a single Q vector is involved in the $\mathbf{Q}\mathbf{K}^T$ operation in the decode stage, leading to underutilization of the pipelined routers in the Q-channel RPUs. 

    The throughput drops sublinearly with the increase of model sizes, which stems from both how models scale and the critical path in the proposed architecture. 
    Model size typically scales along three dimensions: the embedding dimension, the MLP hidden dimension, and the number of layers, whose scaling factors are denoted as $s_e$, $s_h$, and $s_l$, respectively. 
    Under such scaling, the attention and MLP layers approximately increase in parameter count by factors of $(s_e^2)\times$ and $(s_e\cdot s_h)\times$, respectively. 
    For example, when comparing Llama 3.2-1B and Llama 3-8B, it is $s_e=2$, $s_h=1.75$, and $s_l=2$, resulting in an overall model size increase of roughly $\sim$8$\times$ model. 
    However, thanks to the proposed row-wise and column-wise partitioning in both spatial and temporal mapping, the critical path for operations such as broadcast, reduction, and DDMMs is primarily determined by the longest horizontal or vertical communication route. 
    As a result, the critical path scales approximately with $s_e \cdot s_l$ or $s_h \cdot s_l$, instead of the full $s_e \cdot s_h \cdot s_l$ factor, which explains the sublinear drop in throughput. 

    Fig.~\ref{fig-instruction-breakdown} shows the breakdown of clock cycles by instruction along the critical path when processing an attention layer and its corresponding MLPs in Llama 3.2-1B, for both the prefill and decode stages. 
    Thanks to the overlapping of computation and communication, along with the high parallelism of PIM PEs, PIM operations rarely lie on the critical path. 
    Instead, latency is predominantly bottlenecked by data movement and DDMM operations within the IRCUs. 
    To investigate ways to alleviate this bottleneck, we evaluated throughput under different packet bit-widths and IRCU parallelism levels. 
    The resulting trends are presented in Fig.~\ref{fig-throughput-roofline}, which illustrates the trade-offs between communication bandwidth and compute parallelism. 
    The roofline analysis confirms that the configuration used in this work — 64-bit packet width and 16-way parallelism in the IRCU — achieves near-optimal throughput at the performance frontier, without incurring excessive resource overhead.

\section{Related Work}

    \input{tables/table_pim_accelerator_comparison}

\subsection{Parallelism in Machine Learning}

    \subsubsection{LLM in Distributed GPUs}
    Various parallelism strategies on datasets, model weights, context windows, etc, have been adopted for scaling LLMs on GPU clusters~\cite{Meta-2024-llama3.1}. 
    These approaches often rely on high-performance collective communication fabrics such as NVLink to manage synchronization overhead. 
    In contrast, the fine-grained parallelism enabled by tensor partitioning and RPU dataflow optimization in this work is tailored to low-level on-chip interconnects, making it more suitable for domain-specific and energy-constrained applications. 

    \subsubsection{Neural Network Accelerators}
    Neural networks (NNs) have been widely accelerated using spatial architectures, including fully digital designs~\cite{JSSC-2017-Eyeriss, JETCAS-2019-EyerissV2} and PIM-based architectures~\cite{TCAD-2018-NeuSimV1, TCAD-2020-NeuSimV2, ISCAS-2021-DesFramework, JETCAS-2023-BenchmarkMapping}. 
    Much of the analysis in NN accelerators focuses on exploiting parallelism by unrolling the nested loops inherent in convolution operations~\cite{MICRO-2019-MAESTRO, ASPLOS-2020-interstellar}. 
    Compared to attention layers in LLMs, these nested loops tend to be deeper, but the operand matrices are smaller and generally limited to DSMMs rather than the hybrid of DSMMs and DDMMs of LLMs.

\subsection{Emerging Spatial Architecture}

    Spatial architectures distribute modularized memory and computing resources across a spatial array, enabling flexible parallel execution and enhanced data locality. 
    Emerging examples in ML acceleration include coarse-grained reconfigurable arrays (CGRAs)~\cite{DAC-2023-ML-CGRA, ICCAD-2023-Flex} and Cerebras Wafer-Scale Engine (WSE)~\cite{AX-2025-WaferLLM}. 
    In contrast to these existing designs, this work incorporates heterogeneous memory and compute resources within each PIM-NoC macro, providing heterogeneous optimization space for both DSMMs and DDMMs.

\subsection{PIM-based LLM Accelerator}

    A range of customized PIM designs have been proposed for LLM acceleration, as summarized in Table~\ref{table-pim-comparison}. 
    ReTransformer~\cite{ICCAD-2020-retransformer} enhances attention pipelining by decomposing intermediate matrices to improve dataflow efficiency. 
    TranCIM~\cite{ISSCC-2022-trancim} introduces transposable SRAM arrays and implements a coarse-grained pipeline across Q/K/V stacks using a dedicated streaming interconnect. 
    CPSAA~\cite{TCAD-2023-CPSAA} improves the attention pipeline by applying transposition on input activations and supports unstructured dynamic sparsity pruning for DDMMs. 
    HALO~\cite{JETCAS-2024-HALO} leverages a 2.5D integrated architecture that combines PIM-based chiplets for DSMMs and systolic array-based chiplets for DDMMs, optimizing inter-chiplet communication through coarse-grained mapping strategies. 
    H3D-Transformer~\cite{TODAES-2024-H3D} adopts a hybrid-precision approach: low-precision PIM arrays approximate DSMMs, while high-precision digital units refine results to preserve accuracy. 
    While most existing works focus on algorithm-specific optimizations with limited scalability, this paper proposes a scalable architecture that integrates modular compute/memory/communication resources and flexible dataflow to enhance LLM acceleration. 

%% file: tables/table_system_setup.tex
\begin{table}[t]
\caption{System-level hardware configuration}
\begin{center}
\begin{tabular}{cc|cc}

\hline

\textbf{Component} & \textbf{Specs} & \textbf{Component} & \textbf{Specs} \\

\hline
\hline

\multicolumn{4}{c}{\textbf{Architecture level} (for Llama 3.2-1B)} \\

\hline

Tile \# & 64 & Channel \# & 4 per tile \\

RPU \# & 32 per channel & Macro \# & 8 per RPU \\

\hline
\hline






\multicolumn{4}{c}{\textbf{Macro level}} \\

\hline

XB size & 128$\times$128 & XB cell & 8-bit \\

Scratchpad size & 32 KB & Scratchpad width & 16-bit \\

Rout. buf. size & 256 B & Rout. buf. width & 16-bit \\

Packet width & 64-bit & MAC \# & 16 \\

\hline

\end{tabular}
\label{table-hardware-configuration}
\end{center}
\end{table}

%% file: tables/table_hardware_evaluation_7nm.tex
\begin{table}[t]
\caption{Macro-level power and area breakdown}
\begin{center}
\begin{tabular}{c|cc|cc}
\hline

& Power ($\mu W$) & Breakdown & Area ($mm^2$) & Breakdown  \\

\hline
\hline

PIM PE & 32.37~\cite{TCASI-2019-optimizing-mapping} & 15.08\% & 0.0864~\cite{TCASI-2019-optimizing-mapping} & 73.16\% \\

\hline

Scratchpad & 37.80 & 23.53\% & 0.0125 & 10.58\% \\

\hline

Router* & 90.48 & 56.32\% & 0.021 & 17.78\% \\

\hline
\hline

Total & 160.65 & 100\% & 0.1181 & 100\% \\

\hline

\multicolumn{5}{l}{* The digital results are obtained on 45 nm PDK and then scaled to 7 nm. }

\end{tabular}
\label{table-macro-breakdown}
\end{center}
\end{table}

%% file: tables/table_throughput_comparison_hardware.tex
\begin{table}[t]
\caption{Comparison to GPU platforms}
\begin{center}
\begin{tabular}{c|c|c|c|c}

\hline

\multicolumn{2}{c|}{} & Ours & A100 & H100 \\

\hline

\multicolumn{2}{c|}{Frequency (GHz)}  & 1 & 1.4 & 1.7 \\

\hline

Throughput* & Llama 3-8B & 202.25 & 78.36 & 274.26 \\

(tokens/s) & Llama 2-13B & 120.62 & 47.86 & 167.51 \\

\hline

\multicolumn{2}{c|}{Power (W)} & 10.53 & $\sim$300 & $\sim$350 \\

\hline

Energy efficiency & Llama 3-8B & 19.21 & 0.2612 & 0.7836 \\

(tokens/J) & Llama 2-13B & 11.45 & 0.1628 & 0.4786 \\

\hline

\multicolumn{5}{l}{* Tested context window: 1024 input tokens, and 1024 output tokens. }

\end{tabular}
\label{table-gpu-comparison}
\end{center}
\end{table}

%% file: tables/table_pim_accelerator_comparison.tex
\begin{table*}[t]
\caption{Comparison to SOTA PIM-related LLM accelerators}
\begin{center}
\begin{tabular}{c|c|c|c|c}

\hline

 & \textbf{Projection} & \textbf{Attention} & \textbf{Interconnect} & \textbf{Dataflow} \\

\hline

LEAP (\textbf{This work}) & PIM & In-router dataflow accelerator & 2D mesh network-on-chip & Fine-grained parallelism \\

\hline

ReTransformer~\cite{ICCAD-2020-retransformer} & PIM & PIM & Customized on-chip bus & Matrix decomposition + fine-grained pipelining \\

\hline

TranCIM~\cite{ISSCC-2022-trancim} & PIM & PIM & Customized on-chip bus & Coarse-grained pipelining \\

\hline

CPSAA~\cite{TCAD-2023-CPSAA} & Hybrid PIM & Hybrid PIM & Customized on-chip bus & Sparsity-aware \\

\hline

HALO~\cite{JETCAS-2024-HALO} & Hybrid PIM & Systolic array & 2.5D network-on-package & Coarse-grained mapping optimization \\

\hline

H3D-Transformer~\cite{TODAES-2024-H3D} & PIM & PIM + systolic array & 2.5D network-on-package  & Hybrid-precision + coarse-grained pipelining \\

\hline
\end{tabular}
\label{table-pim-comparison}
\end{center}
\end{table*}

%% file: sec-conclusion.tex
\section{Conclusion}

This paper presents an aggregated NoC-PIM architecture for LLM inference acceleration that fully amortizes the computations in the memory and routers. 
A dedicated end-to-end framework orchestrates the model partitioning, mapping, and scheduling, ensuring high resource utilization and parallelism. 
Evaluation results demonstrate substantial improvement in throughput and energy efficiency for the Llama model family compared to the on-shelf GPUs. 
Furthermore, the architecture is highly scalable, accommodating model growth and expanding context windows, with potential for integration with future wafer-scale and network-on-package technologies.